\documentclass[a4paper, two-column, twocolappendix]{aa}
\makeatletter
\renewcommand*\aa@pageof{, page \thepage{} of \pageref*{LastPage}}
\makeatother

\pdfoutput=1
\usepackage{color, xcolor, graphicx, amssymb, mathrsfs, placeins, times, xspace}
\usepackage{deluxetable}
\usepackage{natbib}
\usepackage{appendix}
\usepackage{txfonts}
\usepackage{orcidlink}

\newcommand{\percent}{\ensuremath{\%}}
\newcommand{\Msun}{M_\odot}

\newcommand{\Mtot}{\mathrm{M}_\mathrm{tot}}

\newcommand{\Mbar}{\mathrm{M}_\mathrm{bar}}
\newcommand{\gbar}{\mathrm{g}_\mathrm{bar}}
\newcommand{\gobs}{\mathrm{g}_\mathrm{obs}}
\newcommand{\gdag}{\mathrm{g}_\mathrm{\dagger}}
\newcommand{\gddag}{\mathrm{g}_\mathrm{\ddagger}}
\newcommand{\gM}{\mathrm{g}_{\mathrm{M}}}

\begin{document}

\title{A Distinct Radial Acceleration Relation across Brightest Cluster Galaxies and Galaxy Clusters}

\author{Yong Tian\,\orcidlink{0000-0001-9962-1816}
        \inst{1}
        \and
        Chung-Ming Ko\,\orcidlink{0000-0002-6459-4763}
        \inst{1,2}\fnmsep\thanks{Corresponding author}
        \and
        Pengfei Li\,\orcidlink{0000-0002-6707-2581}
        \inst{3}\fnmsep\thanks{Humboldt fellow.}
        \and
        Stacy McGaugh\,\orcidlink{0000-0002-9762-0980}
        \inst{4}
        \and
        Shemile L. Poblete
        \inst{1}
        }

\institute{Institute of Astronomy, National Central University, Taoyuan 320317, Taiwan
            \and
             Department of Physics and Center for Complex Systems, National Central University, Taoyuan 320317, Taiwan\\
            \email{cmko@astro.ncu.edu.tw}
            \and
            Leibniz-Institute for Astrophysics, An der Sternwarte 16, 14482 Potsdam, Germany
            \and
            Department of Astronomy, Case Western Reserve University, 10900 Euclid Avenue, Cleveland, OH 44106, USA\\
            \email{stacy.mcgaugh@case.edu}
             }

   \date{Received xxx; accepted xxx}

%
%

\abstract
{
Recent studies reveal a radial acceleration relation (RAR) in galaxies, which illustrates a tight empirical correlation connecting the observational acceleration and the baryonic acceleration with a characteristic acceleration scale. However, a distinct RAR has been revealed on BCG-cluster scales with a seventeen times larger acceleration scale by the gravitational lensing effect.
In this work, we systematically explored the acceleration and mass correlations between dynamical and baryonic components in 50 Brightest Cluster Galaxies (BCGs).
To investigate the dynamical RAR in BCGs, we derived their dynamical accelerations from the stellar kinematics using the Jeans equation through Abel inversion and adopted the baryonic mass from the SDSS photometry. We explored the spatially resolved kinematic profiles with the largest integral field spectroscopy (IFS) data mounted by the Mapping Nearby Galaxies at Apache Point Observatory (MaNGA) survey.
Our results demonstrate that the dynamical RAR in BCGs is consistent with the lensing RAR on BCG-cluster scales as well as a larger acceleration scale.
This finding may imply that BCGs and galaxy clusters have fundamental differences from field galaxies. We also find a mass correlation, but it is less tight than the acceleration correlation.
}


\maketitle

%
%

\section{Introduction}

Tight empirical scaling relations are pivotal in physics and astronomy, particularly in the exploration of new fundamental concepts. 
A prominent example is the dark matter (DM) problem, manifested through the discrepancy between observed gravitational effects (inferred mass) and the mass calculated from luminosity (baryonic matter). 
However, an intriguing perspective arises when examining the empirical acceleration relation in galaxies. This perspective focuses on the discrepancy between the baryonic acceleration $\gbar\equiv\,G\Mbar(<r)/r^{2}$ and the observed acceleration $\gobs$, ascertained through methods like gravitational lensing, rotation curve analysis, or the Jeans equation. 
This inquiry into galactic accelerations uncovers a significant empirical scaling relation, which includes the identification of a characteristic acceleration scale, a key parameter in understanding these dynamics.

Recent discoveries have unveiled a tight empirical radial acceleration relation~\citep[RAR;][]{McGaugh16, Lelli17} in spiral galaxies, 
 providing a fresh perspective on the dark matter problem. 
The correlation can be parameterized between $\gobs$ and $\gbar$ as
\begin{equation}
\label{eq:RAR}
 \gobs=\frac{\gbar}{1-e^{-\sqrt{\gbar/\gdag}}}\,,
\end{equation} 
 which exhibits a characteristic acceleration scale $\gdag=(1.20\pm0.02)\times10^{-10}$m\,s$^{-2}$~\citep{McGaugh16, Lelli17, Li18}.
Additionally, the low acceleration limit ($\gbar\ll\gdag$) recovers 
 the baryonic Tully-Fisher relation~\citep[BTFR;][]{McGaugh00, Verheijen01, McGaugh11, Lelli16, Lelli19}, $v_{\mathrm{f}}^4=G\Mbar\gdag$, 
 which relates the flat rotation velocity $v_{\mathrm{f}}$ and the baryonic mass $\Mbar$.
Subsequent studies of the RAR in elliptical galaxies have yielded consistent results, 
 not only in dynamics~\citep{Lelli17, Chae19, Chae20} but also in gravitational lensing effects~\citep{TK19, Brouwer21}. 
In elliptical galaxies, replacing $v_{\mathrm{f}}$ with the velocity dispersion $\sigma$ 
 yields the baryonic Faber-Jackson relation~\citep[BFJR;][]{Sanders10, FM12}, 
 stating that $\sigma^4\propto G\Mbar\gdag$.
These findings further emphasize the importance of the RAR in understanding the fundamental concepts 
 behind the observed acceleration discrepancies in galaxies.
Coincidentally, the RAR has been predicted by MOdified Newtonian Dynamics~\citep[MOND;][]{Milgrom83} 
 four decades ago.

In addition to galaxies, the ``missing mass'' problem has also been observed 
 in galaxy clusters~\citep{Zwicky33, FM12, Overzier16, Rines16, Umetsu20}, 
  which represent the largest gravitational bound systems in the universe. 
Within the strong gravitational potential of galaxy clusters, 
 the brightest cluster galaxy (BCG) is typically located at the center. 
However, even when accounting for the baryonic mass in the form of X-ray gas, 
 the calculated mass remains insufficient when determining gravity using dynamics or gravitational lensing effects.

Upon examining the acceleration discrepancy on BCG-cluster scales, 
 recent studies have revealed a more significant offset 
 from the RAR in galaxy clusters~\citep{CD20, Tian20, Pradyumna21, Eckert22, Tam23, Liu23, li23}. 
In particular, a tight RAR~\citep{Tian20} has been investigated in 20 massive galaxy clusters 
 from the Cluster Lensing And Supernova survey with Hubble (CLASH) samples, 
 referred to as the CLASH RAR, expressed as 
\begin{equation}\label{eq:CLASHRAR}
 \gobs\simeq\sqrt{\gbar\gddag}\,,
\end{equation}
This relation features a larger acceleration scale, $\gddag=(2.0\pm0.1)\times10^{-9}$m\,s$^{-2}$, 
 and a small lognormal intrinsic scatter of $15^{+3}_{-3}\percent$.
The CLASH RAR implies a parallel baryonic Faber-Jackson relation (BFJR) in galaxy clusters, 
 given by $\sigma^4\propto G\Mbar\gddag$. 
More recent studies confirm this kinematic counterpart as 
 the mass-velocity dispersion relation~\citep[MVDR;][]{Tian21a, Tian21b},
 found in 29 galaxy clusters and 54 BCGs.
The MVDR exhibits a consistent acceleration scale $\gddag=(1.7\pm0.7)\times10^{-9}$m\,s$^{-2}$,
 with the CLASH samples and a small lognormal intrinsic scatter of $10^{+2}_{-1}\percent$.

 In this work, we investigate the dynamical RAR in BCGs to address the second acceleration scale, $\gddag$, 
  on BCG-cluster scales. 
The confirmation of a distinct RAR necessitates a thorough examination of both gravitational mass and baryonic mass. 
To accomplish this, we employ integral field spectroscopy (IFS) for the internal kinematics of BCGs and model photometry for the stellar mass. 
It is intriguing to investigate both acceleration and mass correlations within the same sample. 
Moreover, by comparing dynamical and lensing samples, a universal acceleration scale can be scrutinized in galaxies, particularly for BCGs.

%
%

\section{Data and Methods}\label{sec:Method}

Our primary objective is to investigate the correlation between the dynamical acceleration 
 and the baryonic acceleration in BCGs, as well as their mass correlation. 
While the RAR in 20 CLASH samples has been examined in gravitational lensing~\citep{Tian20}, 
 the RAR of BCGs in dynamics has not been systematically explored. 
Furthermore, we assess the consistency of the dynamical and lensing RAR and
 compare the results between BCGs and clusters. 

Investigating the dynamical RAR requires spatially resolved velocity dispersion profiles and the estimation of baryonic mass for individual galaxies. 
In our BCG samples, we noted that most velocity dispersion profiles within our samples follow linear trends. This observation facilitates a simplified approach to analyzing galactic dynamics through the Jeans equation. Our primary objective is to derive the observed gravitational acceleration $\gobs$ with enhanced accuracy. To this end, we employed Abel's inversion~\citep{BM82, BT08}, an analytic formulation for linear velocity dispersion profiles. This technique, when combined with Bayesian statistics for fitting the velocity dispersion profiles, enables us to efficiently compute $\gobs$ using Abel's inversion of the Jeans equation. On the other hand, $\gbar$ is calculated using the accumulated baryonic mass at the same radius, employing an empirical S{\'e}rsic profile.

  \begin{figure*}[hbt!]
  \centering
  \includegraphics[width=2.0\columnwidth]{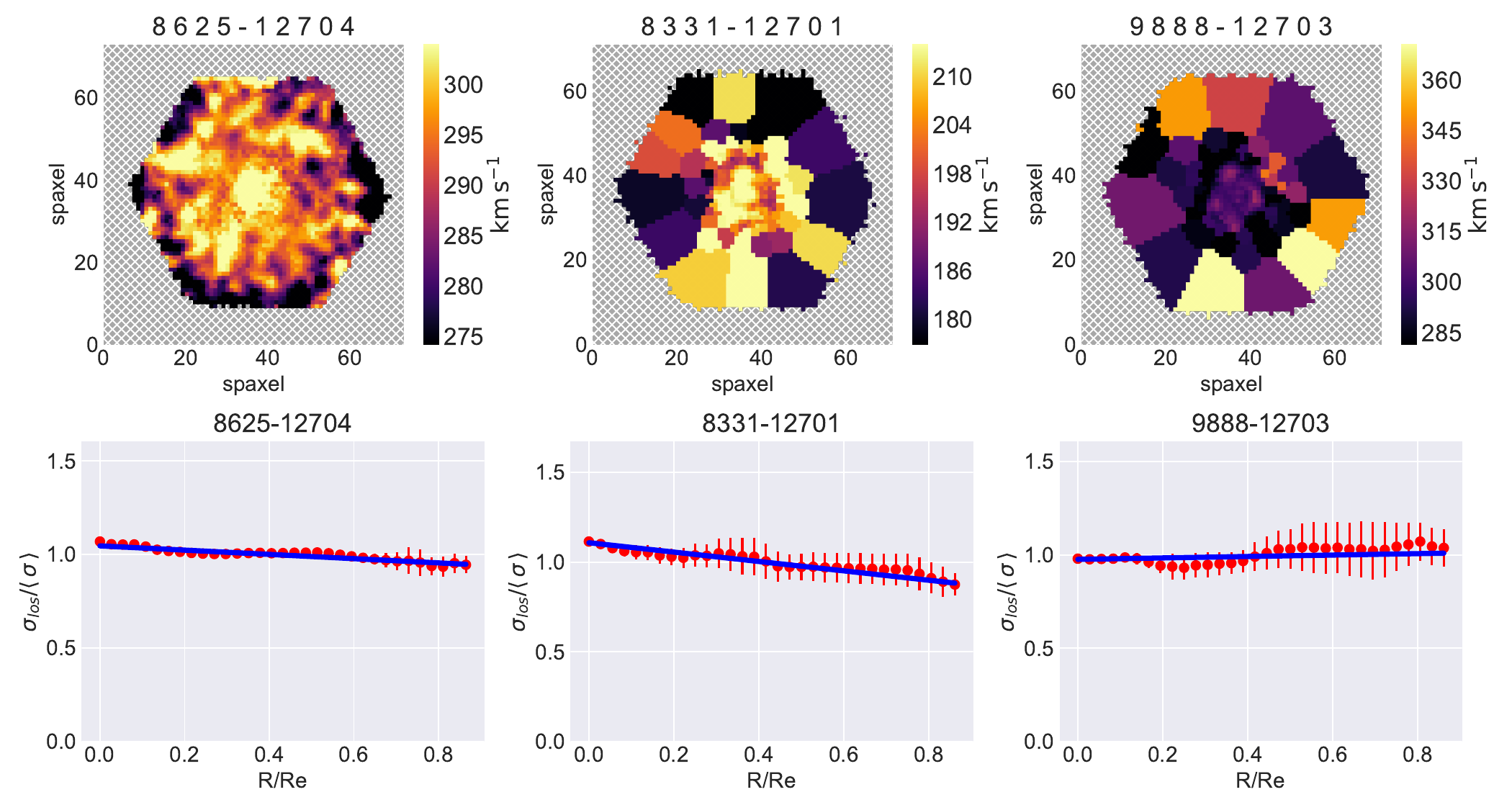}
  \caption{
  \footnotesize
Three examples of MaNGA BCGs with the plateifu of 
 `8554-6102', `8331-12701', and `9888-12703'.
Upper panel: The two-dimensional map plot of \texttt{Spaxel} data for the stellar velocity dispersion. 
Lower panel: velocity dispersion profiles in terms of effective radius.
The red circles represent the los velocity dispersion of concentric circles at different radii with the corresponding error bar.
The solid blue lines represent the linear fit with the ODR MLE method.
  }\label{fig:1}
  \end{figure*}

    \subsection{Data}\label{sec:Data}
Spatially resolved kinematic profiles in galaxies necessitate IFS 
 to measure spectra for hundreds of points within each galaxy. 
At present, MaNGA provides an unprecedented sample of approximately 10,000 nearby galaxies to investigate the internal kinematic structure and composition of gas and stars~\citep{Bundy15}. 
MaNGA is a core program in the fourth-generation Sloan Digital Sky Survey~\citep[SDSS;][]{Law21}.
Also, MaNGA offers one of the most extensive samples of BCGs for IFS studies, presenting an unparalleled resource for in-depth analysis.

BCGs are distinguished not only by their luminosity and mass but also by their unique morphological and kinematic characteristics, which set them apart from typical galaxies. 
Typically located at the centers of galaxy clusters, BCGs often exhibit elliptical or cD galaxy morphologies, characterized by extended, diffuse envelopes indicative of their evolutionary history marked by mergers and accretion within dense cluster environments. 
Kinematically, BCGs generally exhibit significant velocity dispersion with less pronounced rotational features.

In this study, we investigate 50 MaNGA BCGs with complete IFS and photometry data for dynamical analysis, 
 building upon the previous systematic exploration of their kinematics.
The velocity dispersion of MaNGA BCGs has been systematically explored using IFS data~\citep{Tian21a}, 
 so we further investigated these BCG samples for dynamic studies. 
Most of these BCGs were initially obtained from Yang's catalog~\citep{Yang07}, 
 which was developed using a halo-based group finder in the SDSS Data Release 4 (DR4). 
\cite{Hsu21} re-identified them visually in color-magnitude diagrams with the corresponding memberships. 
From their 54 galaxy sample, one BCG `8943-9102' is excluded due to its morphology being identified as a spiral galaxy.
Because BCGs are generally classified as elliptical galaxies, identifying BCGs with spiral galaxy morphology may be considered a misclassification or misidentification.
Additionally, three BCGs, `8613-12705', `9042-3702', and `9044-12703', are excluded due to uncertain stellar mass estimated from model photometry in SDSS. 
Ultimately, we utilized 50 BCGs with complete IFS and photometry data, listed with their plateifu IDs in Table~\ref{tab:BCGs}.

To examine the baryonic mass contribution of BCGs, we carefully considered the total stellar and gas mass.
In our study, the stellar mass within the MaNGA survey is estimated by utilizing photometric data from the SDSS, where galaxy luminosities are converted to stellar masses through a constant mass-to-light ratio~\citep{Law21}. Concurrently, the gas mass is ascertained primarily from the analysis of ionized gas emission lines, which are observed using MaNGA's integral field spectroscopy \citep{Bundy15}.
The gas fraction in our samples accounts for less than $3.2\percent$ of the baryonic mass. 
However, MaNGA primarily studies ionized gas in galaxies using optical spectroscopy through emission lines like H$\alpha$, [NII], [OII], [OIII], and [SII], but it's not designed for observing neutral atomic (HI) or molecular gas (H2)~\citep{Bundy15}. Its focus on optical wavelengths might miss regions with high dust or dominated by neutral or molecular gas, possibly underestimating total galaxy gas content.
Because the RAR should be estimated at the same radius for both $\gobs$ and $\gbar$,
we compute the velocity dispersion measured up to approximately the effective radius and the accumulated baryonic mass based on the measured Sérsic profile at the same radius.
Furthermore, we explored an alternative scenario regarding the possible gradient in the mass-to-light ratio in Appendix~\ref{appendix:A}. This analysis yields a stellar mass variation in the range of approximately 10\% to 20\%.

    \subsection{Velocity Dispersion Profile with Bayesian statistics}\label{sec:Bayesian}
We adopted the line-of-sight (los) one-dimensional velocity dispersion of 50 MaNGA BCGs~\citep{Tian21a} and 
 employed the profile with a linear fit by Bayesian statistics.
\cite{Tian21a} compute the mean los stellar velocity dispersion of each circle relative to their centers,  
 by \texttt{Marvin} developed in \textit{Python}~\citep{Cherinka19}.
Surprisingly, those velocity dispersion profiles demonstrate remarkably flat even at the innermost region for most cases~\citep{Tian21a}.
Instead of the flat profile, we improved the fitting with a linear relation 
 and estimated the error for each data point by implementing a Maximum likelihood Estimation with 
 the orthogonal-distance-regression (ODR) method suggested in \cite{Lelli19, Tian21b, Tian21a}
The linear relation with the ODR method is adopted with the log-likelihood function as
$-2\ln\,\mathcal{L}=\sum_{i}\,\ln{(2\pi\sigma^2_{i})}+\sum_{i}\,\Delta_i^2/\sigma^2_{i}\,,$
 with $\Delta_i^2=(y_i-m\,x_i-b)^2/(m^2+1)\,$, 
 where $i$ runs over all data points, and $\sigma_i$ includes the
 observational uncertainties $(\sigma_{x_i}, \sigma_{y_i})$ and
    the lognormal intrinsic scatter $\sigma_\mathrm{int}$ by
\begin{equation}\label{eq:sigma}
     \sigma^2_{i}=\frac{m^2\sigma^2_{x_i}}{m^2+1}+\frac{\sigma^2_{y_i}}{m^2+1}+\sigma_\mathrm{int}^2\,.
\end{equation}

We model the one-dimensional velocity dispersion by two variables of $y\equiv\sigma/\langle\sigma\rangle$ and $x\equiv\mathrm{R}/\mathrm{R_{e}}$, which is normalized to mean velocity dispersion $\langle\sigma\rangle$ and effective radius $\mathrm{R_{e}}$.
In our samples, only the uncertainty of velocity dispersion is obtained for fitting.
The slopes and the intercepts with the ODR MLE are presented in Table~\ref{tab:BCGs} and 
 illustrated for three examples in Fig.~\ref{fig:1}. 

    \subsection{Dynamical Mass Inferred by Abel Inversion}
In pressure-supported systems such as BCGs, 
 the dynamics in equilibrium are governed by the Jeans equation in spherical coordinates~\citep{BM82, BT08}:
\begin{equation}\label{eq:Jeans}
    \frac{d(\rho\sigma^2_{r})}{dr}+\frac{2\beta}{r}\rho\sigma^2_{r}=-\rho\,\gobs\,.
\end{equation}
where $\beta=1-(\sigma^2_{t}/\sigma^2_{r})$ represent the anisotropy parameter.
For simplicity, we consider the isotropic case as $\beta=0$ and express the projected velocity dispersion $\sigma_{I}(R)$ in the form of an Abel integral equation with its inverse~\citep{BM82, Mamon05, BT08}:
\begin{equation}\label{eq:projected_vd}
    \rho\sigma^2_{r}=\frac{-\Upsilon}{\pi}\int^{\infty}_{r}\frac{dI(R)\sigma^2_{I}(R)}{dR}\frac{dR}{\sqrt{R^2-r^2}}\,,
\end{equation}
where $I(R)$ represents the surface density and $\Upsilon$ denotes the mass-to-light ratio, depending on $R$ in general..
Then, we can conduct $\gobs$ through Abel inversion, deducing from Eq.~(\ref{eq:Jeans}) and Eq.~(\ref{eq:projected_vd}) expressed as~\citep{BM82, Mamon05, BT08}
\begin{equation}\label{eq:Abel}
   \gobs(r) = \frac{-\sigma^2_{r}}{\Upsilon}\frac{d\Upsilon}{dr} + \frac{r\Upsilon}{\pi\rho}\int^{\infty}_{r}\frac{d}{dR}\left(\frac{dI(R)\sigma^2_{I}(R)}{RdR}\right)\frac{dR}{\sqrt{R^2-r^2}}\,.
\end{equation} 
In this study, we model $\sigma_{I}(R)$ using a linear relation in the velocity dispersion profile due to mostly flat velocity dispersion profiles in our MaNGA BCG samples, see Fig.~\ref{fig:1}.
Consequently, the total mass in Newtonian dynamics is defined as $\Mtot(<r)=r^2\gobs(r)/G$. 
Additionally, we estimate the deviation for the anisotropic models in Appendix~\ref{appendix:B}, resulting in at most 6$\percent$ (or a scatter of 0.02 dex) for $\gobs$.

%
%

\section{Results}\label{sec:Results}

Our primary objective is to investigate the dynamical mass and the RAR in MaNGA BCGs and compare the results with the lensing RAR observed in CLASH BCGs and clusters. 
The observational acceleration is directly computed at the last data point using Abel's inversion applied to the velocity dispersion profiles. 
On the other hand, the baryonic acceleration is estimated from the accumulated stellar mass, which is modeled with a S{\'e}rsic profile.
Both accelerations are measured independently and presented on different axes. 
Moreover, we employ Bayesian statistics to assess the tightness of the correlation and examine their residuals with four galactic and cluster properties. 
Additionally, we illustrate the relationship between dynamical and baryonic mass for comparative purposes.

\subsection{Radial Acceleration Relation \& Mass Correlation}

\begin{figure*}[htb!]
  \centering
  \includegraphics[width=1.95\columnwidth]{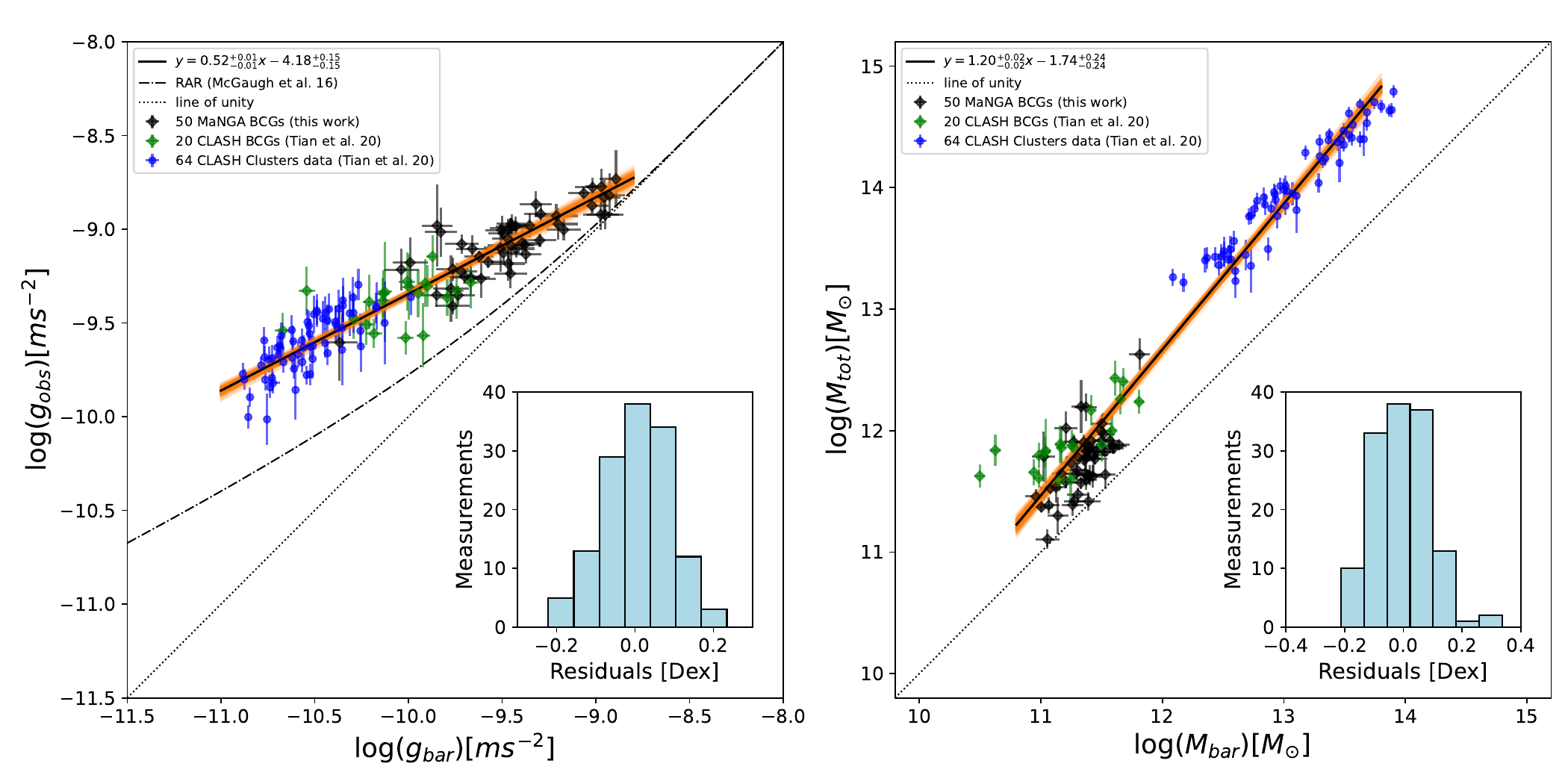}
  \caption{
\footnotesize
The RAR and mass correlation of both BCGs and clusters are presented. 
\textit{Left panel}: Black and green diamonds represent 50 MaNGA and 20 CLASH BCGs, while blue circles indicate 64 CLASH galaxy cluster data.
The black solid line illustrates the resulting RAR of all samples: $\log(\gobs/\mathrm{ms}^{-2})=0.52^{+0.01}_{-0.01}\log(\gbar/\mathrm{ms}^{-2})-4.18^{+0.15}_{-0.15}$.
The orange shaded area illustrates the $1\sigma$ error of the best fit with the ODR MCMC method. 
The inset panel demonstrates the histograms of the orthogonal residuals of a whole sample (blue). 
For comparison, the galactic RAR is depicted by the dot-dashed line, while the dotted line represents the line of unity.
\textit{Right panel}: The black solid line represents the mass correlation of all samples: $\log(\Mtot/\Msun)=1.20^{+0.02}_{-0.02}\log(\Mbar/\Msun)-1.74^{+0.24}_{-0.24}$.
All symbols are the same as those in the left panel.
  }\label{fig:2}
\end{figure*}

We implemented an ODR Markov Chain Monte Carlo (MCMC) analysis, to explore the linear correlation evident in the RAR. 
Utilizing the \textit{python} package~\citep[emcee;][]{corner, emcee19}, we conducted an ODR MCMC analysis 
 to estimate the slope $m$, intercept $b$, and intrinsic scatter $\sigma_{\rm int}$
 with two variables of $y\equiv\log(\gobs/\mathrm{ms}^{-2})$ and $x\equiv\log(\gbar/\mathrm{ms}^{-2})$. 
We employed non-informative flat priors for the slope and intercept within the range of $[-100, 100]$, and for the intrinsic scatter with $\ln(\sigma_{\rm int})\,\in[-10,10]$. 
The findings from our ODR MCMC analysis are illustrated in Fig.~\ref{fig:2} for various samples.

Our sample, consisting of 50 MaNGA BCGs, showed a linear correlation with a slope of $y=0.58^{+0.05}_{-0.04}x-3.62^{+0.43}_{-0.42}$, which dominated the higher acceleration region. 
When we combined MaNGA BCGs with the lensing result of 20 CLASH BCGs, the linear correlation presents a shallow slope of $y=(0.54\pm0.03)x-(4.01\pm0.29)$, which extended to a lower acceleration region.
Finally, we performed a complete fit including all 50 MaNGA BCGs, 20 CLASH BCGs, and 64 clusters data, which resulted in the following equation:
\begin{equation}\label{eq:BCG_RAR}
 \log\left(\frac{\gobs}{\mathrm{ms}^{-2}}\right)=0.52^{+0.01}_{-0.01}\log\left(\frac{\gbar}{\mathrm{ms}^{-2}}\right)-4.19^{+0.15}_{-0.15}\,,
\end{equation}
 with a tiny uncertainty lognormal intrinsic scatter of $(4.9\pm0.9)\percent$, corresponding to 0.02 dex.
The related triangle diagrams of the regression parameters are presented in Appendix~\ref{appendix:C}.

In further analysis, we employed the same vertical MCMC method used in the CLASH RAR study~\citep{Tian20}. 
This approach yielded a shallow slope represented by the relation $y=(0.50\pm0.01)x-(4.30\pm0.15)$, 
 along with a similar uncertainty in the lognormal intrinsic scatter of $(5.6\pm1.0)\percent$, corresponding to 0.02 dex. 
While the fitting result may vary slightly depending on the methods employed, 
 the differences remain minor and are consistent with their respective uncertainties.

We subsequently analyzed the mass correlation in a logarithmic diagram 
 between the accumulated total mass $\Mtot$ and baryonic mass $\Mbar$, 
 as shown in the right panel of Fig.~\ref{fig:2}. 
Using the ODR MCMC method for the entire sample to establish a linear correlation, 
 the result yielded $\log(\Mtot/\Msun)=(1.20\pm0.02)\log(\Mbar/\Msun)-(1.74\pm0.24)$ 
 with a larger uncertainty in the lognormal intrinsic scatter of $(19\pm2)\percent$. 
Unlike the consistency observed in acceleration correlation, 
 CLASH BCGs and MaNGA BCGs dominate the same mass range but exhibit a larger scatter 
 in mass correlation when combined.

\subsection{Residuals}
\begin{figure*}[htb!]
\centering
\includegraphics[width=1.95\columnwidth]{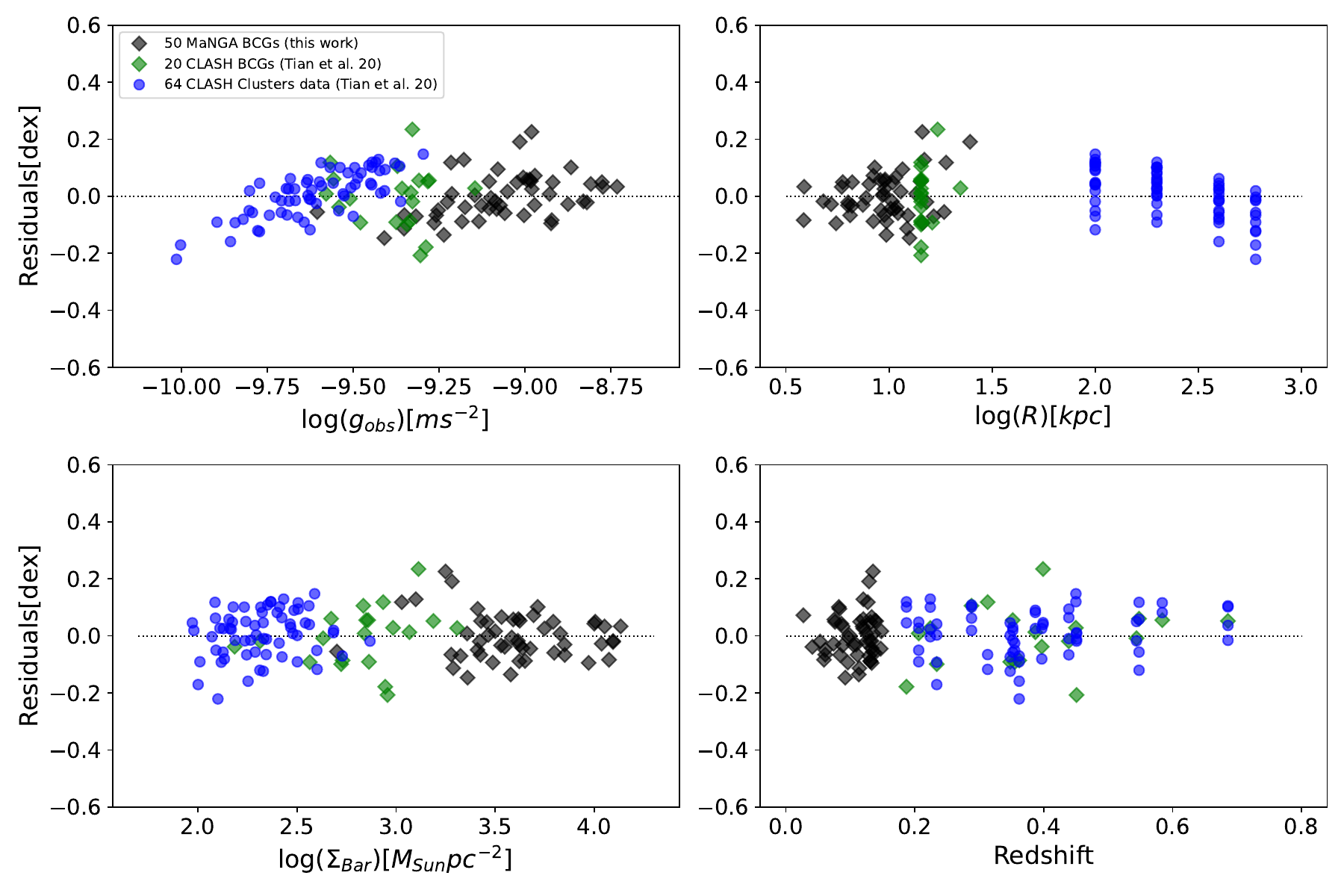}
\caption{
\footnotesize
The orthogonal residuals against four global quantities of galaxies and galaxy clusters: 
 the observational acceleration $\gobs$ (upper-left), radius R (upper-right), baryonic mass surface density $\Sigma_{\mathrm{Bar}}$ (lower-left), and redshift (lower-right). 
The plot displays the residuals for 50 MaNGA BCGs (black diamonds), 20 CLASH BCGs (green diamonds), and 64 data points from CLASH clusters (blue circles), with the dashed line representing zero difference. 
}\label{fig:4}
\end{figure*}

To investigate correlations of the residuals, we compute the orthogonal distance between Eq.~(\ref{eq:BCG_RAR}) and each individual data against four global quantities of BCGs and clusters: the observational acceleration, radius, baryonic mass surface density, and redshift, see Fig.~\ref{fig:4}. 
The residuals of all samples are distributed within a tiny range from $-0.22$ to $0.23$ dex.
No significant correlations were observed in the residuals diagram, except for a slight correlation
of cluster data concerning the outer radius.

\subsection{Mimicking Baryonic Tully-Fisher Relation}
\begin{figure}[htb!]
\centering
\includegraphics[width=1.0\columnwidth]{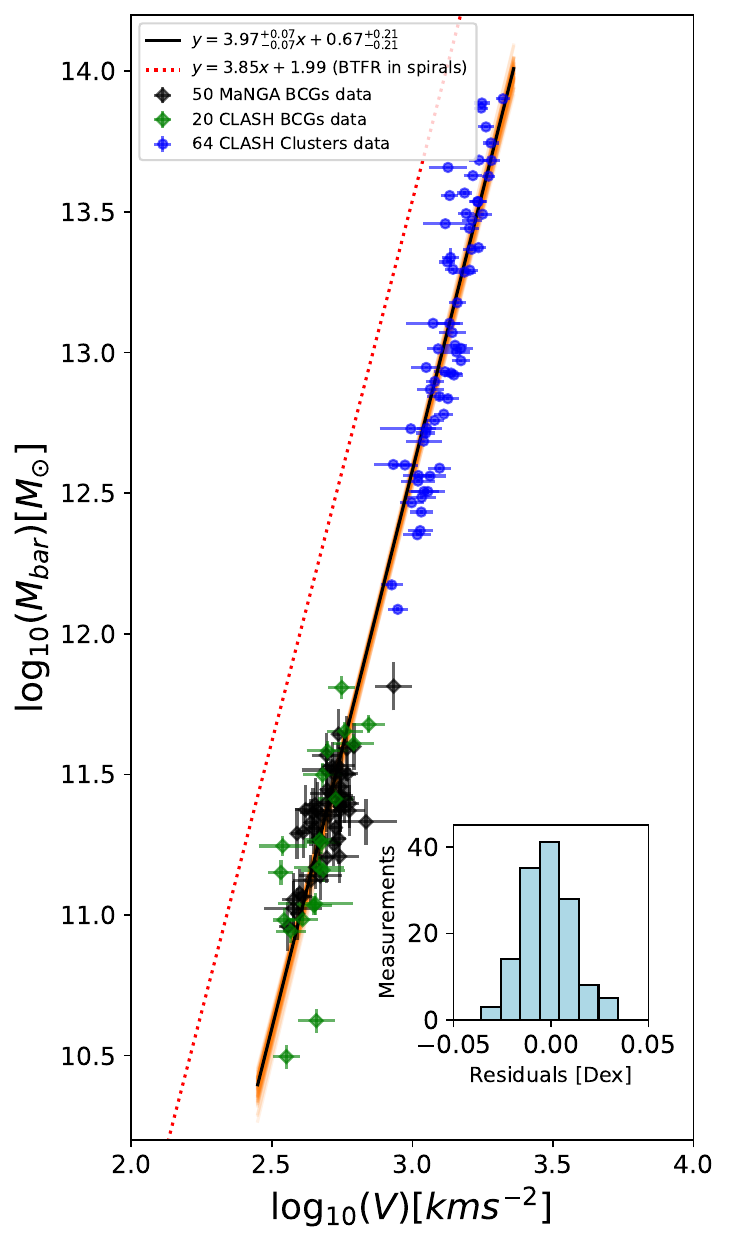}
\caption{
\footnotesize
The black solid line illustrates the fitting BTFR of all samples: $\log(\Mbar/\Msun)=(3.97\pm0.07)\log(V/\mathrm{ms}^{-1})-(0.67\pm0.21)$.
The orange shaded area illustrates the $1\sigma$ error of the best fit with the ODR MCMC method. 
The inset panel demonstrates the histograms of the orthogonal residuals of a whole sample (blue). 
For comparison, the galactic BTFR~\citep{Lelli19} is depicted by the red dotted line.
The plot displays the residuals for 50 MaNGA BCGs (black diamonds), 20 CLASH BCGs (green diamonds), and 64 data points from CLASH clusters (blue circles).
}\label{fig:3}
\end{figure}

Although the approach may seem conceptual, it is enlightening to analyze the BTFR as the kinematic analog of the tight dynamical scaling relation in this context. 
Defining the circular velocity as $V\equiv\sqrt{\gobs\,r}$, we can correspond $V$ using the total acceleration of both BCG and cluster samples. 
The BTFR can be investigated with both $V$ and the baryonic mass $\Mbar$. 
One of the main benefits of examining the BTFR is the expansion of the dynamic range of the distinct RAR concerning the baryonic mass. 
This range spans approximately 3.5 orders of magnitude in our samples, as illustrated in Fig.~\ref{fig:3}.

Because our samples demonstrate a linear correlation, we adopted the ORD MCMC again for analysis.
The resulting BTFR presents a tight relation with uncertainty in the lognormal intrinsic scatter of $(11\pm2)\percent$, equivalent to $0.05$ dex, and can be expressed as
\begin{equation}\label{eq:BTFR}
 \log\left(\frac{\Mbar}{\Msun}\right)=3.97^{+0.07}_{-0.07}\log\left(\frac{V}{\mathrm{kms}^{-1}}\right)+0.67^{+0.21}_{-0.21}\,.
\end{equation}
The slope of this relation closely aligns with that observed in spiral galaxies~\citep{Lelli19}, 
 being nearly parallel with a value close to four.
However, a deviation is noted in the intercept, suggesting a larger acceleration scale. 
By assuming a fixed slope of four for $\gddag$,
 the relation simplifies to $\log(\Mbar/\Msun)=4\log(V/\mathrm{kms}^{-1})+(0.57\pm0.02)$, 
 maintaining the intrinsic scatter of $0.05$ dex.
Because Equation~(\ref{eq:CLASHRAR}) induced a parallel BTFR, represented as $V^4=G\Mbar\gddag$,
 we are able to determine $\gddag=(2.0\pm0.1)\times10^{-9}$ms$^{-2}$. 
This value aligns consistently with the $\gddag$ obtained from the distinct RAR.

%
%

\section{Discussions and Summary}\label{sec:Discussions}

In this study, we report four key discoveries that significantly enhance our understanding 
 of the RAR and its implications for BCGs and clusters. 
Our main findings are as follows:
(1) We identified a linear correlation between MaNGA BCGs, CLASH BCGs, and clusters in RAR; 
(2) The acceleration scale $\gddag$ was found to be valid across a range spanning over two orders of magnitude in baryonic acceleration, suggesting the RAR to hold on even larger scales than previously considered;
(3) The distribution of residuals within a narrow range highlights a tight correlation between the dynamical and baryonic components in BCGs and clusters; and
(4) There is no significant accumulated mass correlation on BCG-cluster scales.

The distinct RAR on BCG-cluster scales provides a fresh perspective 
 on the residual missing mass of MOND in galaxy clusters~\citep{Sanders99, Sanders03,  Angus08, Milgrom08, Angus10, FM12}, 
 recasting it as an acceleration-dependent residual mass issue. 
Although MOND successfully explained the missing mass problem in galactic systems~\citep{BZ22}, it has been reported that additional mass, such as missing baryons \citep{Sanders99, Sanders03, Milgrom08} or sterile neutrinos \citep{Angus10, FM12}, is required in galaxy clusters, a phenomenon known as the residual missing mass.
When examining the RAR within the context of the MOND framework, 
 we can compute the residual missing mass by comparing it with two RARs, as described below.
Here, $\gbar$ represents the baryonic acceleration estimated by the baryonic mass in our samples, 
 while $\gM$ denotes the MONDian mass necessary to reproduce a distinct RAR. 
One can define a missing mass ratio $Q=\gM/\gbar=M_{M}/\Mbar$ in MOND when considering the same observational acceleration. 
To explore the possibility of compensating for the distinct RAR with missing mass, we connect the RAR as fitted in \cite{McGaugh16} to the same observed acceleration $\gobs$ by
\begin{equation}\label{eq:missing baryon}
    \sqrt{\gbar\gddag} \approx \frac{\gM}{1 - e^{-\sqrt{\gM/\gdag}}}\,.
\end{equation}
Using Eq.~(\ref{eq:missing baryon}), we can compute the factor $Q$ for a given $\gbar$. For example, with a median logarithm of baryonic acceleration in 50 BCGs at $\log(\gbar) = -9.5$, we find $Q = 2.2$. However, the value of $Q$ exhibits significant variability with $\gbar$: for the highest acceleration in our MaNGA BCG samples, $\log(\gbar) = -8.9$, $Q$ decreases to $1.2$, while for the lowest acceleration, $\log(\gbar) = -10.4$, $Q$ increases to $4.9$. This variability indicates that a systematic constant offset in the mass-to-light ratio is insufficient to address the discrepancy. Consequently, it appears that the residual missing mass is correlated with the baryonic acceleration $\gbar$.

In this study, we also evaluated the mass-to-light ratio $\Upsilon$ to validate the choice of using total stellar mass calculated from SDSS model photometry.
For 50 MaNGA BCGs, we calculated $\Upsilon$ across the five SDSS bands: u, g, r, i, and z. 
The average values for these bands are $(\Upsilon_\mathbf{u}, \Upsilon_\mathbf{g}, \Upsilon_\mathbf{r}, \Upsilon_\mathbf{i}, \Upsilon_\mathbf{z})=(9.6, 5.5, 3.1, 2.3, 1.9)$. 
Our results indicate that the potential underestimation of the stellar mass is a serious concern. 
The residual missing mass would be other forms rather than the stellar mass, such as the underestimation of gas mass or sterile neutrinos.

Besides the residual mass on BCG-cluster scales, other possibility can be further investigated within the framework of relativistic MOND theories.
Interestingly, given the consistency between our MaNGA BCGs and the CLASH samples measured by gravitational lensing, relativistic MOND offers a variety of results that extend beyond the standard MOND formulation. 
One particularly intriguing interpretation involves the acceleration scale being contingent on the depth of the potential well~\citep[eMOND;][]{ZF12, HZ17}.
In BCGs and galaxy clusters, eMOND implied a stronger observational acceleration that deviates from the RAR suggested by the initial MOND.
Moreover, recent advancements in relativistic MOND theories~\citep{SZ21, Verwayen23} also suggest the feasibility of an enhancement in galaxy clusters.

Investigating the mass consistency in MaNGA BCGs reveals implications 
 for the dark matter distribution in galaxy clusters.
Some MaNGA BCGs exhibit a consistent mass between dynamical and baryonic mass, 
 suggesting insufficient dark matter within one effective radius. 
This finding poses a challenge to the merger model in the CDM paradigm, 
 where one would expect a significant amount of dark matter at the center of galaxy clusters due to dynamical friction.
To fully comprehend this issue, further analysis of these specific BCG samples is necessary, especially compared with computer simulations such as TNG~\citep{Springel18, Nelson19a, Nelson19b}, EAGLE~\citep{Schaye15, Crain15}, and BAHAMAS~\citep{McCarthy17}, etc.

Investigating the RAR in the context of BCGs and galaxy clusters is key to understanding the dark matter problem, particularly in relation to the residual missing mass. MOND, which accurately predicts the RAR's slope of 0.5, indicates a correlation between the residual missing mass and baryonic acceleration, warranting further exploration. Additionally, the RAR's application to BCG-cluster scales is crucial for assessing different dark matter theories, where it notably challenges the self-interacting dark matter model, especially in light of the BAHAMAS simulation results \citep{Tam23}. Although our current dynamical RAR studies in BCGs would benefit from improved stellar mass estimation methods that are model-independent and enhanced dynamical mass evaluations using numerical models \citep{li23}, these initial findings broaden the scope of baryonic acceleration research and establish a foundation for more detailed future investigations.

\begin{acknowledgements}
We sincerely appreciate the referee's constructive suggestions and valuable comments, which have significantly contributed to enhancing the clarity and overall quality of our work.
YT is supported by the Taiwan National Science and Technology Council NSTC 110-2112-M-008-015-MY3. 
CMK and SLP are supported by the Taiwan NSTC 111-2112-M-008-013 and NSTC 112-2112-M-008-032.
PL is supported by the Alexander von Humboldt Foundation.
SSM is supported in part by NASA ADAP grant 80NSSC19k0570 and also acknowledges support from NSF PHY-1911909.
\end{acknowledgements}

\bibliographystyle{aa}
\bibliography{reference}

\begin{deluxetable}{lcclccccccc}
\centering
\tablecaption{\label{tab:BCGs}Properties and Results of 50 MaNGA BCGs}
    \tablehead{
         \multicolumn{1}{c}{plateifu} &
         \multicolumn{1}{c}{$z$\tablenotemark{(a)}} &
         \multicolumn{1}{c}{$\mathrm{N}_\mathrm{S\acute{e}rsic}$} &
         \multicolumn{1}{c}{$\mathrm{R}_\mathrm{eff}$} &
         \multicolumn{1}{c}{$\mathrm{r}_\mathrm{last}$\tablenotemark{(b)}} &
         \multicolumn{1}{c}{$m$\tablenotemark{(c)}} &
         \multicolumn{1}{c}{$b$\tablenotemark{(c)}} &
         \multicolumn{1}{c}{$\log(\Mbar)$\tablenotemark{(d)}} &
         \multicolumn{1}{c}{$\langle\sigma_\mathrm{los}\rangle$\tablenotemark{(e)}}  &
         \multicolumn{1}{c}{$\log(\gbar)$} &
         \multicolumn{1}{c}{$\log(\gobs)$} \\
         \colhead{}
         & \colhead{} & \colhead{} &
         \multicolumn{1}{c}{[kpc]} &
         \multicolumn{1}{c}{$\mathrm{R}_\mathrm{eff}$} &
         & \colhead{} &
         \multicolumn{1}{c}{[$\Msun$]} &
         \multicolumn{1}{c}{[km/s]} &
         \multicolumn{1}{c}{[m/s$^2$]} &
         \multicolumn{1}{c}{[m/s$^2$]}
    }
\startdata	
8625-12704$^{*}$	&	0.027	&	4.6	&	9.7	&	0.86	&	-0.114	&	1.046	&	11.541 $\pm\,0.077$	&	293	$\pm\,11$	&	-9.45 $\pm\,0.08$	&	-8.97 $\pm\,0.03$	\\
9181-12702$^{*}$	&	0.041	&	4.9	&	13.0	&	0.84	&	-0.154	&	1.039	&	11.622 $\pm\,0.086$	&	265 $\pm\,10$	&	-9.57 $\pm\,0.09$	&	-9.17 $\pm\,0.03$ \\
9492-9101	&	0.053	&	4.0	&	18.0	&	0.84	&	-0.124	&	1.048	&	11.785 $\pm\,0.085$	&	293 $\pm\,12$	&	-9.72 $\pm\,0.09$	&	-9.23 $\pm\,0.04$	\\
8258-3703	&	0.059	&	4.5	&	5.3	&	0.73	&	-0.401	&	1.138	&	11.246 $\pm\,0.096$	&	192 $\pm\,18$	&	-8.97 $\pm\,0.10$	&	-8.92 $\pm\,0.08$	\\
8331-12701	&	0.061	&	6.0	&	11.0	&	0.86	&	-0.261	&	1.108	&	11.231 $\pm\,0.086$	&	197 $\pm\,14$	&	-9.85 $\pm\,0.09$	&	-9.35 $\pm\,0.06$	\\
8600-12703$^{*}$	&	0.061	&	6.0	&	10.9	&	0.84	&	0.029	&	0.985	&	11.332 $\pm\,0.086$	&	219 $\pm\,10$	&	-9.70 $\pm\,0.09$	&	-9.25 $\pm\,0.04$	\\
8977-3703$^{*}$	&	0.074	&	4.1	&	7.2	&	0.73	&	0.040	&	0.976	&	11.475 $\pm\,0.086$	&	248 $\pm\,12$	&	-9.02 $\pm\,0.09$	&	-8.87 $\pm\,0.04$	\\
8591-3704	&	0.075	&	6.0	&	11.9	&	0.76	&	-0.075	&	1.010	&	11.506 $\pm\,0.106$	&	283 $\pm\,10$	&	-9.46 $\pm\,0.11$	&	-8.99 $\pm\,0.03$	\\
8591-6102	&	0.076	&	6.0	&	8.2	&	0.81	&	-0.235	&	1.097	&	11.435	$\pm\,0.089$	&	262 $\pm\,16$	&	-9.29 $\pm\,0.09$	&	-8.92 $\pm\,0.05$	\\
8335-6103	&	0.082	&	6.0	&	11.5	&	0.74	&	0.127	&	0.934	&	11.575 $\pm\,0.086$	&	321 $\pm\,25$	&	-9.32 $\pm\,0.09$	&	-8.87 $\pm\,0.07$	\\
9888-12703	&	0.083	&	6.0	&	13.5	&	0.86	&	0.038	&	0.976	&	11.542 $\pm\,0.089$	&	307 $\pm\,18$	&	-9.72 $\pm\,0.09$	&	-9.08 $\pm\,0.05$	\\
9043-3704$^{*}$	&	0.084	&	6.0	&	8.9	&	0.73	&	-0.254	&	1.091	&	11.473 $\pm\,0.091$	&	225 $\pm\,15$	&	-9.17 $\pm\,0.09$	&	-9.00 $\pm\,0.06$	\\
8613-6102	&	0.086	&	6.0	&	10.3	&	0.78	&	-0.016	&	0.989	&	11.807 $\pm\,0.083$	&	334 $\pm\,13$	&	-9.06 $\pm\,0.08$	&	-8.81 $\pm\,0.03$	\\
9002-3703	&	0.088	&	6.0	&	9.1	&	0.73	&	-0.152	&	1.048	&	11.171 $\pm\,0.092$	&	200 $\pm\,1$	&	-9.49 $\pm\,0.09$	&	-9.13 $\pm\,0.04$	\\
8939-6104	&	0.088	&	6.0	&	5.3	&	0.73	&	-0.682	&	1.123	&	11.309 $\pm\,0.086$	&	267 $\pm\,47$	&	-8.89 $\pm\,0.09$	&	-8.73 $\pm\,0.15$	\\
8455-12703$^{*}$	&	0.092	&	4.7	&	15.0	&	0.84	&	-0.087	&	1.043	&	11.559 $\pm\,0.091$	&	212 $\pm\,11$	&	-9.76 $\pm\,0.09$	&	-9.41 $\pm\,0.05$	\\
9025-9101	&	0.096	&	5.2	&	17.9	&	0.81	&	0.318	&	0.860	&	11.813	$\pm	\,	0.077	$	&	284	$\pm	\,	34	$	&	-9.61	$\pm	\,	0.08	$	&	-9.26	$\pm	\,	0.10	$	\\
8239-6103	&	0.097	&	6.0	&	9.5	&	0.81	&	-0.543	&	1.152	&	11.361	$\pm	\,	0.088	$	&	245	$\pm	\,	36	$	&	-9.51	$\pm	\,	0.09	$	&	-9.10	$\pm	\,	0.13	$	\\
9486-6103	&	0.098	&	6.0	&	11.5	&	0.81	&	-0.030	&	1.014	&	11.401	$\pm	\,	0.078	$	&	243	$\pm	\,	9	$	&	-9.62	$\pm	\,	0.08	$	&	-9.15	$\pm	\,	0.03	$	\\
9000-9101	&	0.105	&	6.0	&	8.8	&	0.72	&	-0.309	&	1.112	&	11.225	$\pm	\,	0.088	$	&	203	$\pm	\,	16	$	&	-9.39	$\pm	\,	0.09	$	&	-9.07	$\pm	\,	0.07	$	\\
8447-3702	&	0.109	&	6.0	&	9.8	&	0.76	&	-0.375	&	1.153	&	11.596	$\pm	\,	0.106	$	&	251	$\pm	\,	24	$	&	-9.20	$\pm	\,	0.11	$	&	-8.97	$\pm	\,	0.08	$	\\
9891-9101	&	0.111	&	5.8	&	15.2	&	0.81	&	0.171	&	0.877	&	11.609	$\pm	\,	0.101	$	&	268	$\pm	\,	29	$	&	-9.67	$\pm	\,	0.10	$	&	-9.26	$\pm	\,	0.09	$	\\
8466-6104	&	0.113	&	6.0	&	13.1	&	0.74	&	-0.350	&	1.129	&	11.552	$\pm	\,	0.087	$	&	211	$\pm	\,	19	$	&	-9.46	$\pm	\,	0.09	$	&	-9.24	$\pm	\,	0.08	$	\\
9891-3701	&	0.114	&	6.0	&	16.8	&	0.73	&	-0.118	&	1.041	&	11.462	$\pm	\,	0.091	$	&	209	$\pm	\,	12	$	&	-9.73	$\pm	\,	0.09	$	&	-9.35	$\pm	\,	0.05	$	\\
8081-3701	&	0.115	&	6.0	&	8.2	&	0.73	&	-0.208	&	1.075	&	11.556	$\pm	\,	0.094	$	&	281	$\pm	\,	16	$	&	-9.02	$\pm	\,	0.09	$	&	-8.78	$\pm	\,	0.05	$	\\
8131-3703	&	0.119	&	6.0	&	8.1	&	0.73	&	-0.362	&	1.133	&	11.592	$\pm	\,	0.084	$	&	277	$\pm	\,	30	$	&	-8.97	$\pm	\,	0.08	$	&	-8.77	$\pm	\,	0.09	$	\\
9891-12701	&	0.120	&	6.0	&	18.3	&	0.81	&	0.482	&	0.853	&	11.442	$\pm	\,	0.095	$	&	303	$\pm	\,	47	$	&	-9.99	$\pm	\,	0.10	$	&	-9.18	$\pm	\,	0.13	$	\\
9085-6102	&	0.120	&	2.9	&	12.8	&	0.74	&	0.344	&	0.890	&	11.702	$\pm	\,	0.092	$	&	315	$\pm	\,	23	$	&	-9.35	$\pm	\,	0.09	$	&	-8.98	$\pm	\,	0.06	$	\\
9506-6103$^{*}$	&	0.120	&	6.0	&	12.3	&	0.78	&	-0.087	&	1.025	&	11.578	$\pm	\,	0.087	$	&	293	$\pm	\,	22	$	&	-9.45	$\pm	\,	0.09	$	&	-8.99	$\pm	\,	0.07	$	\\
8943-3704	&	0.124	&	6.0	&	8.7	&	0.73	&	-0.519	&	1.195	&	11.697	$\pm	\,	0.082	$	&	269	$\pm	\,	37	$	&	-8.93	$\pm	\,	0.08	$	&	-8.82	$\pm	\,	0.12	$	\\
9490-9102	&	0.125	&	5.0	&	16.1	&	0.84	&	0.035	&	0.972	&	11.619	$\pm	\,	0.088	$	&	284	$\pm	\,	20	$	&	-9.76	$\pm	\,	0.09	$	&	-9.21	$\pm	\,	0.06	$	\\
9043-9101	&	0.127	&	6.0	&	24.2	&	0.78	&	0.615	&	0.785	&	11.582	$\pm	\,	0.092	$	&	338	$\pm	\,	44	$	&	-10.04	$\pm	\,	0.09	$	&	-9.22	$\pm	\,	0.11	$	\\
8725-12704	&	0.129	&	6.0	&	29.5	&	0.84	&	0.655	&	0.770	&	12.068	$\pm	\,	0.086	$	&	499	$\pm	\,	76	$	&	-9.83	$\pm	\,	0.09	$	&	-9.01	$\pm	\,	0.13	$	\\
8989-12704	&	0.129	&	5.3	&	20.3	&	0.81	&	0.083	&	0.872	&	11.758	$\pm	\,	0.094	$	&	300	$\pm	\,	61	$	&	-9.77	$\pm	\,	0.09	$	&	-9.32	$\pm	\,	0.18	$	\\
8989-12703	&	0.130	&	6.0	&	13.2	&	0.78	&	0.491	&	0.832	&	11.644	$\pm	\,	0.124	$	&	284	$\pm	\,	48	$	&	-9.45	$\pm	\,	0.12	$	&	-9.09	$\pm	\,	0.15	$	\\
8728-3703	&	0.131	&	6.0	&	12.0	&	0.81	&	-0.360	&	1.142	&	11.596	$\pm	\,	0.09	$	&	234	$\pm	\,	25	$	&	-9.47	$\pm	\,	0.09	$	&	-9.18	$\pm	\,	0.09	$	\\
9865-12703	&	0.131	&	6.0	&	11.5	&	0.73	&	0.053	&	1.018	&	11.498	$\pm	\,	0.083	$	&	218	$\pm	\,	12	$	&	-9.37	$\pm	\,	0.08	$	&	-9.13	$\pm	\,	0.05	$	\\
8717-1901	&	0.131	&	6.0	&	16.1	&	0.65	&	-0.013	&	0.995	&	11.493	$\pm	\,	0.081	$	&	272	$\pm	\,	8	$	&	-9.49	$\pm	\,	0.08	$	&	-9.02	$\pm	\,	0.03	$	\\
8991-6102	&	0.133	&	6.0	&	13.1	&	0.78	&	-0.083	&	1.024	&	11.654	$\pm	\,	0.099	$	&	264	$\pm	\,	15	$	&	-9.43	$\pm	\,	0.10	$	&	-9.10	$\pm	\,	0.05	$	\\
8555-3702	&	0.133	&	6.0	&	6.3	&	0.76	&	-0.339	&	1.137	&	11.457	$\pm	\,	0.089	$	&	238	$\pm	\,	25	$	&	-8.96	$\pm	\,	0.09	$	&	-8.83	$\pm	\,	0.09	$	\\
8554-6103	&	0.133	&	6.0	&	14.5	&	0.74	&	0.001	&	0.974	&	11.594	$\pm	\,	0.088	$	&	305	$\pm	\,	27	$	&	-9.50	$\pm	\,	0.09	$	&	-9.00	$\pm	\,	0.08	$	\\
8995-6103	&	0.133	&	6.0	&	9.0	&	0.62	&	-0.315	&	1.086	&	11.458	$\pm	\,	0.096	$	&	213	$\pm	\,	19	$	&	-8.95	$\pm	\,	0.10	$	&	-8.92	$\pm	\,	0.08	$	\\
8720-12705	&	0.135	&	6.0	&	23.6	&	0.78	&	-0.250	&	0.950	&	11.236	$\pm	\,	0.096	$	&	230	$\pm	\,	54	$	&	-10.37	$\pm	\,	0.10	$	&	-9.60	$\pm	\,	0.20	$	\\
8615-12704	&	0.135	&	5.1	&	18.0	&	0.81	&	1.095	&	0.538	&	11.572	$\pm	\,	0.081	$	&	495	$\pm	\,	125	$	&	-9.85	$\pm	\,	0.08	$	&	-8.98	$\pm	\,	0.22	$	\\
8554-6102	&	0.136	&	5.6	&	12.7	&	0.74	&	0.044	&	0.965	&	11.775	$\pm	\,	0.115	$	&	312	$\pm	\,	23	$	&	-9.21	$\pm	\,	0.12	$	&	-8.93	$\pm	\,	0.06	$	\\
8616-12703	&	0.138	&	4.0	&	14.2	&	0.78	&	0.135	&	0.923	&	11.884	$\pm	\,	0.089	$	&	309	$\pm	\,	13	$	&	-9.30	$\pm	\,	0.09	$	&	-9.06	$\pm	\,	0.04	$	\\
8616-3702	&	0.138	&	6.0	&	12.5	&	0.76	&	-0.072	&	1.033	&	11.581	$\pm	\,	0.089	$	&	287	$\pm	\,	17	$	&	-9.43	$\pm	\,	0.09	$	&	-8.98	$\pm	\,	0.05	$	\\
8247-9102$^{*}$	&	0.140	&	4.5	&	18.2	&	0.78	&	0.355	&	0.859	&	11.734	$\pm	\,	0.084	$	&	334	$\pm	\,	28	$	&	-9.66	$\pm	\,	0.08	$	&	-9.10	$\pm	\,	0.07	$	\\
9888-6104	&	0.147	&	6.0	&	13.8	&	0.77	&	-0.081	&	1.030	&	11.729	$\pm	\,	0.087	$	&	273	$\pm	\,	11	$	&	-9.38	$\pm	\,	0.09	$	&	-9.08	$\pm	\,	0.03	$	\\
8725-6104	&	0.148	&	6.0	&	17.1	&	0.67	&	0.140	&	0.933	&	11.614	$\pm	\,	0.083	$	&	287	$\pm	\,	35	$	&	-9.47	$\pm	\,	0.08	$	&	-9.05	$\pm	\,	0.11$	
\enddata
\tablecomments{
(a) Redshift from MaNGA Pipe3D;
(b) The last data point in terms of effective radius; 
(c) The slope $m$ and intercept $b$ fitted in the normalized velocity dispersion profile;
(d) The baryonic mass including total stellar mass estimated by model photometry in SDSS DR15 and the measured gas mass in MaNGA marked with $\mathrm{*}$ on the plateifu ID;
(e) The average los velocity dispersion from MaNGA IFS.}
\end{deluxetable} 
\clearpage

\clearpage
\begin{appendix}
\FloatBarrier
\section{The Gradient in the Mass-to-Light Ratio}\label{appendix:A}
To refine our analysis, we incorporated the gradient in the mass-to-light ratio, $\Upsilon_{*}$, 
 within the inner regions of BCGs. Commonly, analyses employing SDSS model photometry operate under the assumption of a constant mass-to-light ratio, $\Upsilon$. 
However, taking into consideration the variation of $\Upsilon_{*}$ within BCGs can have implications. 
We therefore utilized the mass-to-light gradient (Salpeter-Chabrier model) as presented in \cite{Bernardi18}, 
 expressed as $\Upsilon_{*}(\mathrm{R}) = \Upsilon(1+1.29-3.33\mathrm{R}/\mathrm{R}_\mathrm{e})$,
 valid for $\mathrm{R}/\mathrm{R}_\mathrm{e}\leq1.29/3.33$, with the ratio remaining constant thereafter. 
This modification led to an increase in the stellar mass of approximately 10$\percent$ to 20$\percent$ across our sample. 
We conducted a separate analysis of this effect to illustrate the nuances it introduces to our findings.

Upon meticulous examination of the mass-to-light ratio gradient, our observations indicated an absence of any pronounced enhancement in the dynamical acceleration within our BCG sample. This can be attributed to the fact that the mass-to-light ratio tends to stabilize as a constant when $\mathrm{R}/\mathrm{R}_\mathrm{e}$ is sufficiently large. Delving into the Jeans equation, it becomes apparent that the gravitational acceleration $\gobs$ at a given radius $\mathrm{R}$ is contingent upon densities and velocity dispersions beyond that radius. Consequently, for $\gobs$ in the vicinity of $1\mathrm{R}_\mathrm{e}$, variations in the mass-to-light ratio within $\mathrm{R}<0.39\mathrm{R}_\mathrm{e}$ have a negligible influence.

\begin{figure}[!htb]
\centering
\includegraphics[width=1.0\columnwidth]{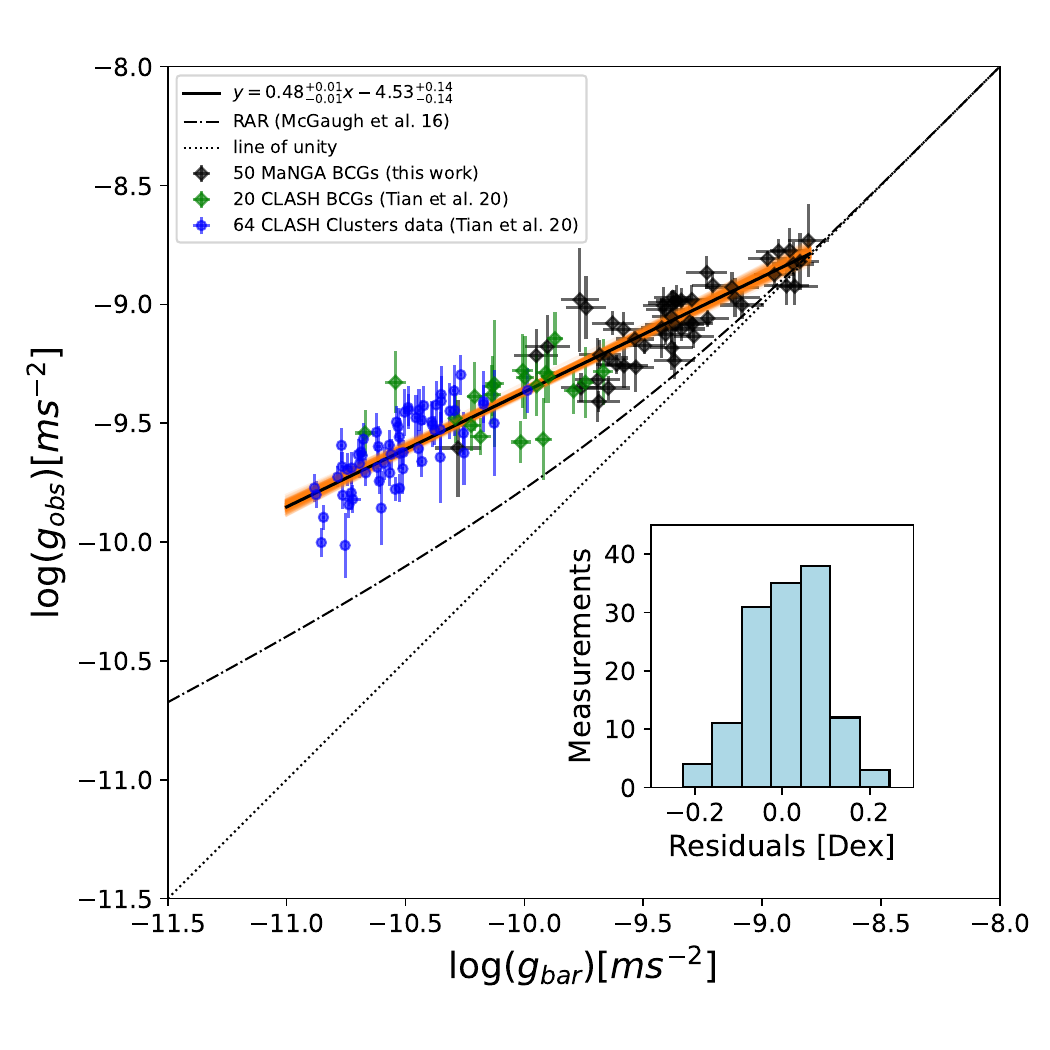}
\caption{
\footnotesize
The RAR of both BCGs and clusters is presented when considering the mass-to-light ratio gradient for 50 MaNGA BCGs. 
All symbols are the same as those in Fig.~\ref{fig:2}.
The black solid line illustrates the resulting RAR of all samples: $\log(\gobs/\mathrm{ms}^{-2})=0.48^{+0.01}_{-0.01}\log(\gbar/\mathrm{ms}^{-2})-4.53^{+0.14}_{-0.14}$. 
For comparison, the galactic RAR is depicted by the dot-dashed line, while the dotted line represents the line of unity.
}\label{fig:A1}
\end{figure}

In analyzing the mass-to-light ratio gradient, 
 we also examined the RAR to understand the influence of an increased baryonic acceleration in our sample. 
This enhancement principally appears as a horizontal shift in logarithmic space of around 0.04 to 0.08 dex. 
Employing ODR with MCMC, we found that this shift gives rise to a shallower slope, 
 represented by the relation $y = 0.48^{+0.01}_{-0.01}x - 4.53^{+0.14}_{-0.14}$, see Fig.~\ref{fig:A1}.
It is noteworthy that this enhancement leads to a small difference between the accumulative dynamical mass and stellar mass in some of the BCG samples, which poses challenges due to the implied deficiency in dark matter.

\section{Scatters estimated by the Anisotropic Parameter}\label{appendix:B}

In our exploration of potential sources of uncertainty, we diligently assessed the influence of anisotropy. 
We adopted the anisotropic model given by $\beta = 0.5\mathrm{r}/(\mathrm{r} + \mathrm{r}_{\mathrm{a}})$
as introduced by \cite{Mamon05}. Additionally, to represent our BCG samples, we utilized a S$\acute{e}$rsic profile with S$\acute{e}$rsic indices $n$ spanning from 2 to 6.

For systems exhibiting anisotropy, we aimed to emulate a nearly flat projected velocity dispersion (within 2$\percent$) to mirror the observed VD profiles of our BCG samples. 
While the observed VD profiles remain flat up to one effective radius in observations, we made
an assumption of this flatness extending up to two effective radii. 
Using the parameter $\mathrm{r}_{\mathrm{a}} = 0.18 \mathrm{r}_{\mathrm{v}} = 14 \mathrm{R}_{\mathrm{e}}$ for $\beta=0.03$ at one effective radius $\mathrm{R}_{\mathrm{e}}$ recommended by \cite{Mamon05}, where $\mathrm{r}_{\mathrm{v}}$ represents the virial radius, the resulting $\gobs$ deviates from the isotropic counterpart by less than 1.2$\percent$ (or a scatter of 0.005 dex) within one effective radius for all S$\acute{e}$rsic indices in the range $2 \leq n \leq 6$. For an extreme case with $\mathrm{r}_{\mathrm{a}} = 1.4 \mathrm{R}_{\mathrm{e}}$ for $\beta=0.21$ at $1\mathrm{R}_{\mathrm{e}}$ \citep{Mamon05, Dekel05}, the $\gobs$ deviates by at most 6$\percent$ (or a scatter of 0.025 dex) within the same radius for all such $n$ values.

Considering alternative anisotropic models, such as 
$\beta = 1.3\mathrm{r}_{\mathrm{a}}\mathrm{r}/(\mathrm{r}^2 + \mathrm{r}_{\mathrm{a}}^2)$
from \citep{Carlberg97, Colin00, Mamon05}, and using $\mathrm{r}_{\mathrm{a}} = 2 \mathrm{r}_{\mathrm{v}} \approx 156\mathrm{R}_{\mathrm{e}}$ for $\beta=0.008$ at $1\mathrm{R}_{\mathrm{e}}$~\citep{Mamon05}, the derived $\gobs$ lies within a 0.3$\percent$ deviation (or 0.001 dex scatter) for all S$\acute{e}$rsic indices in the given range, when measured within one effective radius. The uncertainties stemming from anisotropy closely match the error bars from the measured velocity dispersion. As a result, our RAR calculations under the isotropy assumption emerge as remarkably robust.

\section{Triangle Diagrams of the MCMC Method}\label{appendix:C}
\begin{figure*}[htb!]
\centering
\includegraphics[width=2.0\columnwidth]{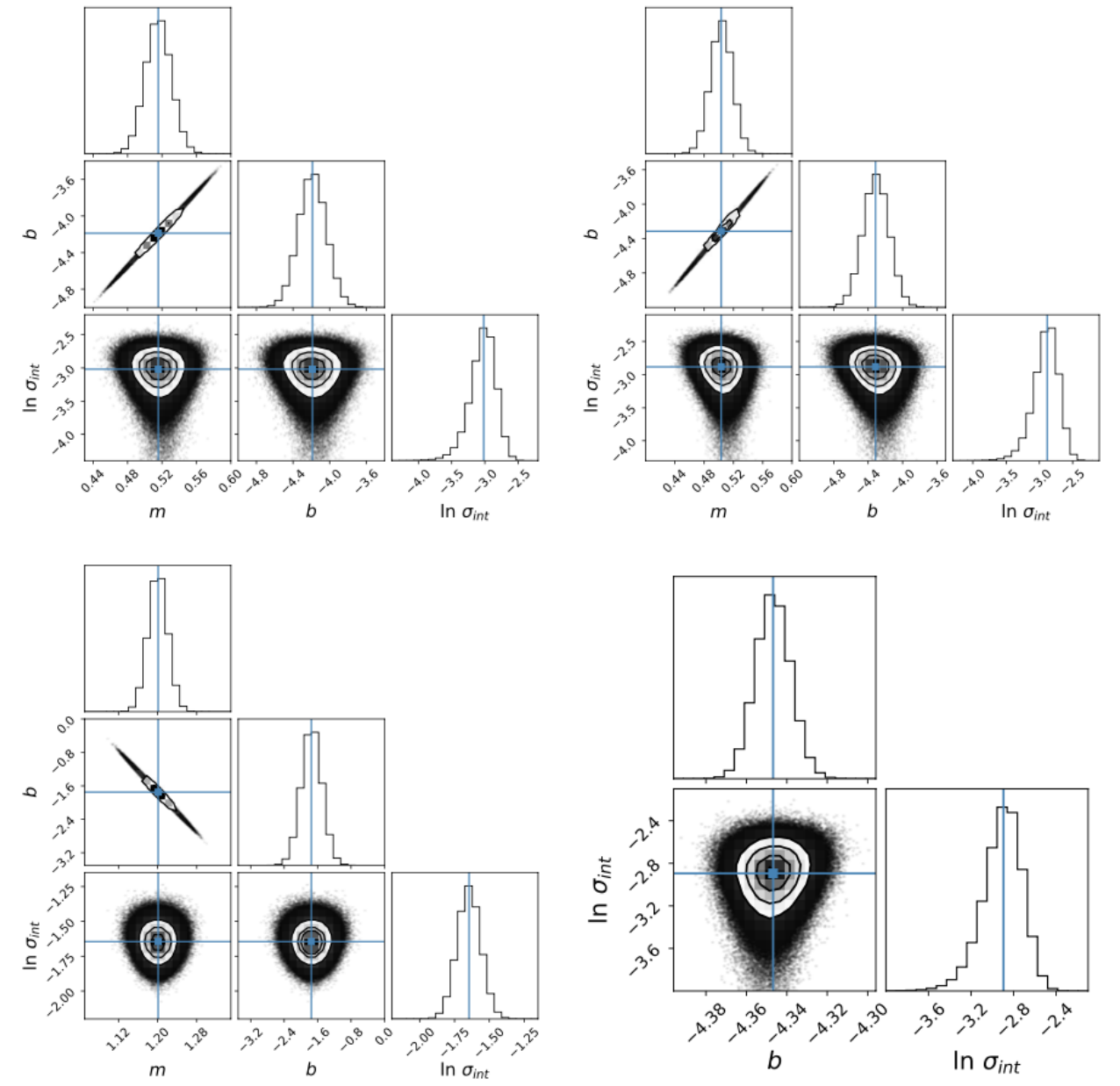}
\caption{
\footnotesize
The MCMC method in 70 BCGs and 64 CLASH cluster data.
\textit{Upper-right panel}: Triangle diagrams of the regression parameters in the ODR MCMC method for the RAR, $y=0.52^{+0.01}_{-0.01}x-4.19^{+0.15}_{-0.15}$, with marginalized one-dimensional (histograms) and two-dimensional posterior distributions.
Black contours represent $1\sigma$ and $2\sigma$ confidence regions.
\textit{Upper-left panel}: Triangle diagrams of the regression parameters in the vertical MCMC method for the RAR, $y=0.50^{+0.01}_{-0.01}x-4.30^{+0.15}_{-0.15}$.
\textit{Lower-right panel}: Triangle diagrams of the regression parameters in the ODR MCMC method for the mass correlation, $y=1.20^{+0.02}_{-0.02}x-1.74^{+0.24}_{-0.24}$.
\textit{Lower-left panel}: Triangle diagrams of the regression parameters in the ODR MCMC method for the RAR with a fixed slope $m=0.5$.}
\label{fig:C1}
\end{figure*}

In our analysis, we employed the MCMC method to investigate the tight correlations. The triangle diagrams depicting the regression parameters are presented in Fig.~\ref{fig:C1} for four distinct scenarios: (1) the ODR MCMC for the RAR; (2) the vertical MCMC for the RAR; (3) the ODR MCMC for the mass correlation; and (4) the ODR MCMC for the RAR with a fixed slope $m=0.5$.

\end{appendix}

\end{document}